\documentstyle[multicol,aps,prl]{revtex}
\begin{document}
\input{psfig}
\draft
\title
{Measurement of Newton's Constant Using a Torsion Balance\\
with Angular Acceleration Feedback}
\author{Jens H. Gundlach and Stephen M. Merkowitz}
\address
{Department of Physics, Nuclear Physics Laboratory,
University of Washington, Seattle, Washington 98195}
\date{\today}
\maketitle
\begin{abstract}
We measured Newton's gravitational constant $G$ using a new torsion balance method.
Our technique greatly reduces several sources of uncertainty 
compared to previous measurements:
(1) it is insensitive to anelastic torsion fiber properties;
(2) a flat plate pendulum minimizes the sensitivity due to the pendulum density distribution;
(3) continuous attractor rotation reduces background noise.
We obtain
$G=(6.674215\pm0.000092)\times10^{-11}$~m$^{3}$~kg$^{-1}$~s$^{-2}$; the Earth's mass is, therefore, $M_{\oplus}=(5.972245\pm0.000082)\times10^{24}$~kg and the Sun's mass
is $M_{\odot}=(1.988435\pm0.000027)\times10^{30}$~kg.
\end{abstract}
\date{\today}
\maketitle
\pacs{PACS numbers: 06.20.Jr, 04.80.-y, 06.30.Gv, 97.10.Nf, 91.10}
\begin{multicols}{2} \global\columnwidth20.5pc
\multicolsep=8pt plus 4 pt minus 3 pt 

The gravitational constant $G$ along with Planck's constant $\hbar$ and the speed of light~$c$, is one of the most fundamental and universal constants in nature. Unlike most other physical constants, the value of $G$ is not precisely known due to the weakness and non-shieldability of gravity. Since the first laboratory measurement by Cavendish over 200 years ago, the reduction in uncertainty in $G$ has been only a factor of about ten per century.
To make matters worse, measurements over the last ten years~\cite{all_meas,bagley} all have larger error-bars than the value of Luther and Towler \cite{luther} on which the 1986 accepted value\cite{codata} was based.
In addition, some of these recent measurements lie far outside the $\pm$128~ppm uncertainty of the 1986 accepted value, bringing this accuracy of $G$ into question. Recognizing this situation, the currently recommended value is $G=(6.673\pm0.010)\times10^{-11}$~m$^{3}$~kg$^{-1}$~s$^{-2}$\cite{codata98}, corresponding to an uncertainty of $\pm$1500~ppm.

We developed a new technique \cite{gundlach1996} that offers several substantial advantages over previous methods.
At the heart of the apparatus is a torsion balance placed on a turntable located between a set of attractor spheres. 
The turntable is first rotated at a constant rate so that the pendulum experiences a sinusoidal torque due to the
gravitational interaction with the attractor masses.
A feedback is then turned on that changes the rotation rate so as to minimize the torsion fiber twist. 
The resulting angular acceleration of the turntable, which is now equal to the gravitational angular acceleration
of the pendulum \cite{rose}, is determined from the second time-derivative of the turntable angle readout.
Since the torsion fiber does not experience any appreciable deflection, this technique is independent
of many torsion fiber properties including anelasticity, which may have led to a bias\cite{kuroda1,bagley,kuroda2} in previous measurements.

The gravitational angular acceleration $\alpha$ of a torsion pendulum expanded in spherical
multipole moments is:
\begin{equation}
\alpha(\phi)=
-\frac{4\pi G}{I}\sum_{l=2}^{\infty}\frac{1}{2l + 1}
\sum_{m=-l}^{+l} m q_{lm} Q_{lm} e^{i m\phi}~,
\label{eqn: alpha}
\end{equation}
where $q_{lm}$ are the multipole moments of the pendulum and the $Q_{lm}$multipole fields\cite{yu} of the external mass distribution, $\phi$ is the turntable angle with respect to the mass distribution, and $I$ is the pendulum moment of inertia about the torsion fiber.\
Since the series converges rapidly, the biggest term in Eq.~\ref{eqn: alpha} is the one with $l=m=2$:
\begin{equation}
\alpha_{22}(\phi)= -\frac{16\pi}{5} G \frac {q_{22}}{I} Q_{22} 
\sin 2\phi~.
\end{equation}
For an ideal, infinitely-thin two-dimensional vertical plate the quotient $q_{22}/I$ is independent of the pendulum mass distribution, $\rho(\vec{r}_{p})$, i.e. mass, size, shape etc.:
\begin{equation}
\frac{q_{22}}{I}
=\frac{\int{}{}\rho(\vec{r}_{p})\ Y_{22}(\theta_{p},\phi_{p})
\ r_{p}^{2}\ d^{3}r_{p}}
{\int{}{}\rho(\vec{r}_{p})\ \sin^{2}\theta_{p}\ r_{p}^{2}\ d^{3}r_{p}}
\stackrel{2D}{\longrightarrow}
\sqrt{\frac{15}{32\pi}}
\end{equation}
and 
\begin{equation}
\alpha(\phi)\approx\alpha_{22}(\phi)= -\sqrt{\frac{24 \pi}{5}}\ G\ Q_{22} \sin 2\phi~.
\end{equation}
The flat pendulum geometry presents a significant advance over previous torsion balance measurements, where the
biggest contribution to the systematic uncertainty was due to the pendulum's mass distribution.
A small and easily calculable modification can be applied to the ideal ratio for a thin rectangular plate with finite thickness $t$ and width $w$:
\begin{equation}
\frac {q_{22}}{I}= \frac{w^2-t^2}{w^2+t^2} \sqrt{\frac{15}{32 \pi}}~.
\end{equation}

All $l>2, m=2$ terms in Eq.~\ref{eqn: alpha} also contribute to the signal $\sin(2\phi)$.
However, these contributions are small since the expansion in Eq.~\ref{eqn: alpha} converges as $(R_{pend}/R_{attr})^{l-2}$, where $R_{pend}$ and $R_{attr}$
are the characteristic radii of the pendulum and the source.
In addition, all $l=$~odd pendulum moments and attractor fields are removed by designing the pendulum ``up-down" symmetric.
The $l=4, m=2$ contribution is also eliminated by design: the $q_{42}$ of a rectangular plate pendulum with height $h$ vanishes by choosing
$h^{2}= \frac{3}{10}(w^{2}+t^{2})$, and the $Q_{42}$ vanishes by using two spheres on each side with their centers spaced vertically by $z=\sqrt{2/3}\rho$, where $\rho$ is the radial distance from the pendulum axis.
The lowest $l>2, m=2$ contribution which does not vanish by design is the small and easily calculable $\alpha_{62}$:
\begin{equation}
\frac{\alpha_{62}}{\alpha_{22}}= \frac{99}{7683200} 
\frac{213(w^4+t^4)+626w^2t^2}{\rho^4}~.
\end{equation}
Table I contains the numeric values of the $\alpha_{62}$ and $\alpha_{82}$ corrections.
We checked the accuracy of our multipole analysis with a full angular acceleration calculation using direct numeric integration.

The attractor spheres are located on a separate coaxial turntable which is rotated with angular velocity
$\omega_{a}(t)=\omega_{d}+\omega_{i}(t)$, where $\omega_{i}(t)$ is the angular velocity of the
torsion balance turntable. The difference angular velocity, $\omega_{d}=\dot{\phi}$, is held constant. Rotation of the attractor masses allows us to cleanly remove gravitational interactions due to the
environment. Furthermore, we are able to set the signal, $\sin(2\omega_{d}t)$, at a relatively
high frequency to suppress the 1/$f$-noise characteristic of the torsion balance and the gravitational background.   

A schematic of the apparatus is shown in Fig.~1. The torsion balance turntable consisted of an air-bearing, a precision angle encoder, and an eddy-current motor.
The torsion pendulum was located in an aluminum vacuum chamber and was surrounded by a $\mu$-metal shield.
The pendulum was hung from a 41.5~cm long, 17~$\mu$m diameter tungsten-fiber, which was attached to a swing damper.
The pendulum was a 1.506~mm thick, 76~mm wide, and 41.6~mm tall Pyrex-glass plate with a thin gold-coating.
The small pendulum deflection angle was sensed with an autocollimator using four reflections off the
pendulum plate.

The centers of the spheres were located at $\rho$=16.76~cm on three stainless steel seats that were embedded in two cast-aluminum shelves. The shelves were supported by a turntable made from a steel bearing.
The attractor spheres were machined from the same selected stock of ultrasonically tested \#316 stainless steel. Their average diameter was 124.89~mm and their mass was $\approx$8.140~kg \cite{qcss}. A pressure-dependent air-density correction was applied to the $G$-measurements.
The apparatus was located in the former cyclotron cave in the Nuclear Physics Lab at the University of Washington on a massive platform 3.5 m above the floor.
The partially underground room was temperature stabilized. Temperature drifts and fluctuations were typically $<0.05$~K/day.
The instrument itself was in a passive thermal enclosure made of Styrofoam. 
The apparatus-temperature was monitored and a correction was applied to compensate for the thermal expansion of the attractor mass assembly during the measurement.

A digital signal processor (DSP) recorded the data and controlled the experiment. The DSP's loop frequency (2.5~KHz)
was used for the timing, which was derived from a temperature-controlled quartz oscillator which was calibrated with a GPS receiver.
The data were averaged by the DSP over exactly one second and uploaded to the host PC.

We recorded six data runs, each approximately three days long. A typical data segment is shown in Fig 2. After each data run the spheres were moved to locations differing by $90^{\circ}$ in azimuthal angle on the attractor turntable shelves. In addition, the spheres from the upper shelf were placed on the lower shelf and the orientation of the spheres were changed in order to average over density fluctuations and non-sphericity. The position of the spheres were measured before and after each data run. We then repeated the entire measurement cycle using four different spheres.

Our largest systematic uncertainty was due to the attractor mass distance measurement. We used a specially fabricated micrometer tool made primarily of Invar for the horizontal distance measurement between the spheres.  Before and after each distance measurement the tool was compared to an Invar ball bar standard that was calibrated at NIST to within 0.2~$\mu$m\cite{nist}. The vertical spacing between the spheres was measured by inserting a small gauge block that was $\approx$10-20~$\mu$m thinner than the gap between the sphere surfaces. We inferred the spacing by optically measuring the angle through which the gauge block could be tilted.
The temperature of the attractor mass was recorded during the distance measurements.

For most of the data $\omega_{i}$ was set to $\approx$5.3~mrad/s and $\omega_{d}$ was set to 20.01015~mrad/s so that the signal frequency occurred at $\approx$6.37~mHz. The rotation frequencies were chosen to be incommensurate. 
We tested a wide variety of $\omega_{i}$ and $\omega_{d}$ and found the results
independent of these angular velocities.

The lab-fixed horizontal magnetic field at the apparatus was measured to be $\approx$100 mG. We ran tests by exaggerating the field at the center of the apparatus to $\approx$5~G and at the location of the spheres to $\approx$100~G.
The observed acceleration difference due to the exaggerated fields was $(6\pm8)\times10^{-12}$~rad/s$^{2}$.
Therefore, a 0.6~ppm error was attributed to magnetic accelerations. 

To investigate the rotating temperature gradient sensitivity, two 5W-heaters were installed on the turntable. These rotating heaters exaggerated the normally observed temperature variation at $2\omega_{d}$ by greater than a factor of 250.
The acceleration difference amplitude with the heaters activated was $|\Delta\alpha|=(21\pm 22)\times10^{-12}$~rad/s$^{2}$.
Therefore the rotating temperature gradient coupled acceleration was $<2\times10^{-13}$~rad/s$^{2}$ (0.4~ppm).

The torsion balance turntable angle was numerically differentiated twice to yield angular
acceleration. All the data were divided into segments comprised of twenty sinusoidal
cycles and fitted using a least squares method. The fitting function included the signal and its harmonics, the room background and its harmonics, an offset and a linear drift. The statistical error was derived from the scatter of the individual fit values.

Table 1 lists the corrections that were applied to the data. Since the DSP averaged over $\tau=1$~s intervals the true amplitudes are attenuated by a small amount: $\frac{\sin(\omega_{d}\tau)}{\omega_{d}\tau}$.
Also the numerical derivative of $\sin(2\omega_{d}t)$ is smaller than the analytic derivative by
$\frac{\sin(\omega_{d}\Delta t)}{\omega_{d}\Delta t}$, where $\Delta t$~$=10$~s was the chosen time increment. Since this correction was substantial we tested a wide variety of increments and found that our results were independent of the choice of $\Delta t$.
We verified the accuracy of the data analysis by simulating the data numerically using various levels of drifts, 1/$f$-noise, and gravitational room background. 

The overall linearity and insensitivity to the signals at $4\omega_{d}$ was tested using two pairs of spheres separated by $45^{\circ}$ on each side. These 8-sphere results agreed with the regular 4-sphere results within the statistical errors.

Due to the finite gain of the acceleration feedback loop a small twist remained in the fiber.
The twist angle at $2\omega_{d}$ was essentially unresolved so that a 0.35~ppm error was used instead of a correction.

We combined the results from the different attractor mass configurations into pairs that averaged out accelerations due to the attractor turntable itself and the combination of three pairs yielded our $G$-value (Fig.~3). After combining the two $G$-values obtained with different spheres our value for the gravitational constant is

\begin{equation}
G=(6.674215\pm0.000092)\times10^{-11}~\mathrm{m}^{3}~\mathrm{kg}^{-1}~\mathrm{s}^{-2}.
\end{equation}
The systematic and statistical uncertainties are shown in Table 2. The total uncertainty is the quadrature sum of the individual uncertainties. Fig.~4 compares our results to other recent measurements. 

Combining our value of $G$ with the geocentric value $GM_{\oplus}$ \cite{ries} we determine the mass of the Earth to be
\begin{equation}
M_{\oplus}=(5.972245\pm0.000082)\times10^{24}~\mathrm{kg}.
\end{equation}
Likewise, using $GM_{\odot}$ \cite{astronomical}, the mass of the Sun is
\begin{equation}
M_{\odot}=(1.988435\pm0.000027)\times10^{30}~\mathrm{kg}.
\end{equation}

We wish to thank the members of the E\"ot-Wash group and Maximilian Schlosshauer for writing the data simulations. Many thanks go to NIST for the support of this work with precision measurement
grant \#~60NANB7D0053 and to NSF for grant \#~PHY-9602494 to the E\"ot-Wash group.
Further assistance was provided by the Physics Department of the University of Washington, and the Nuclear Physics Laboratory which is supported by the DOE.
\begin{table}
\caption{Summary of correction factors.
}
\begin{tabular}{|lr|}
finite pendulum thickness &  1.0007857\\
pendulum attachment and imperfections &  1.0000433\\
$\alpha_{62}$-correction& 0.9998767\\
$\alpha_{82}$-correction&  0.9999951\\
data averaging ($\tau=1$~s, $\omega_{d}=20$~mrad/s)&1.0000667\\
numeric derivatives ($\Delta t=10$~s, $\omega_{d}=20$~mrad/s)&1.0134544\\
\underline{Total:} &   1.0142322\\
\end{tabular}
\label{tab:corrections}
\end{table}
\begin{table}
\begin{flushleft}
\caption{One $\sigma$ error budget.}
\label{tab: errors}
\begin{tabular}{|lcc|}
quantity & measurement & $\Delta G/G$\\
 & uncertainty & \\ \hline
 & & (ppm) \\ \hline
\underline{Systematic errors:} &  & \\
\hspace{0.25cm} pendulum: & & \\
\hspace{0.5cm} width    & $<20$~$\mu$m   & 0.4 \\
\hspace{0.5cm} thickness \& flatness    & $<4.0$~$\mu$m  & 4.0 \\
\hspace{0.25cm} attractor masses: & & \\
\hspace{0.5cm} diagonal separation & $< 1.0$~$\mu$m & 7.1 \\
\hspace{0.5cm} ball-bar calibration & $< 0.2$~$\mu$m & 1.4 \\
\hspace{0.5cm} vertical separation & $< 1.0$~$\mu$m  & 5.2 \\
\hspace{0.5cm} sphere diameter & $< 1.5$~$\mu$m & 2.6 \\
\hspace{0.5cm} temperature uncertainty & $< 100$~mK& 6.9 \\
\hspace{0.5cm} mass & $<3.0$~mg & 0.4\\
\hspace{0.5cm} air humidity&  & 0.5 \\
\hspace{0.25cm} residual twist angle: & &0.3\\
\hspace{0.25cm} magnetic fields: & &0.6\\
\hspace{0.25cm} rot. temperature grad.: & & 0.4 \\
\hspace{0.25cm} time base: & $<10^{-7}$  &0.1 \\
\hspace{0.25cm} data reduction: & & 2.0\\
 & & \\
\underline{Statistical error:} & & 5.8\\
 & &\\
\underline{Total:} & & 13.7\\ 
\end{tabular}
\end{flushleft}
\end{table}
\begin{figure}[p]
\psfig{figure=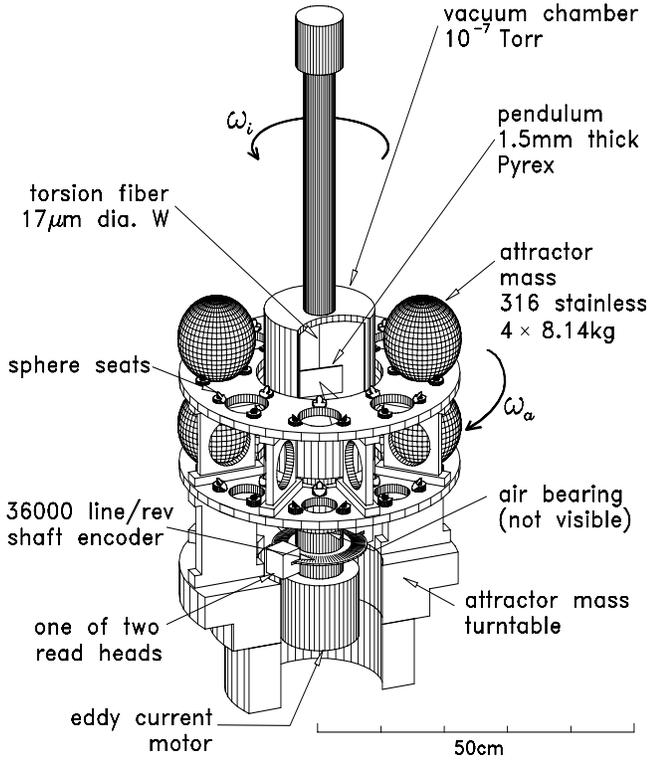,width=8.5cm,height=10.0cm,angle=0}
\caption{
Cut-away view of the apparatus.} 
\label{fig: Bigg-apparatus}
\end{figure}  
\begin{figure}
\psfig{figure=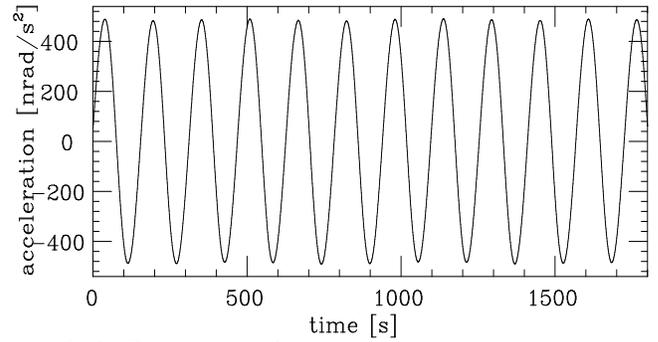,width=8.5cm,height=4.5cm,angle=0}
\caption {Raw acceleration data: a half-hour segment of the twice numerically differentiated turntable angle. The signal frequency was constant and could be freely selected.}
\label{fig: data}
\end{figure}
\begin{figure}[p]
\psfig{figure=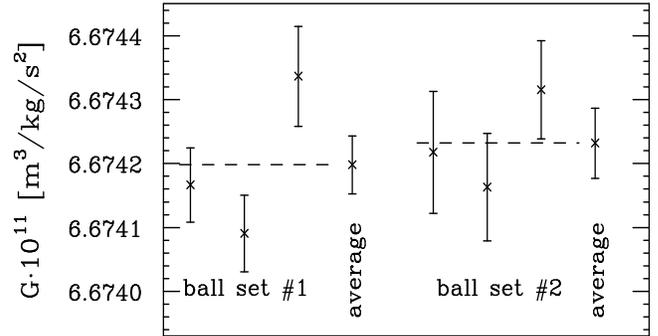,width=8.5cm,height=4.5cm,angle=0}
\caption {The results of two data sets taken with different spheres. Each data point is the combination of a pair of attractor configurations that together eliminate accelerations due to the attractor turntable. Combining the three pairs taken with different sphere orientations optimally reduced
effects from sphere-density and shape imperfections. The displayed uncertainties are statistical only.}
\label{fig:pair_configs}
\end{figure}
\begin{figure}[p]
\psfig{figure=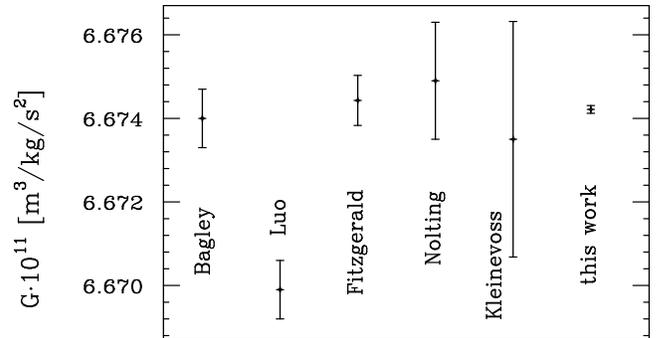,width=8.5cm,height=4.5cm,angle=0}
\caption {Comparison to other measurements\protect\cite{all_meas,bagley} published after 1995 and with
$\Delta G/G<1000$~ppm.}
\label{fig:results}
\end{figure}
\end{multicols}
\end{document}